\documentclass[pre,amsmath,amssymb,nofootinbib,twocolumn, showpacs, superscriptaddress,10pt]{revtex4}

\usepackage[english]{babel}
\usepackage[utf8]{inputenc}
\usepackage[T1]{fontenc}
\usepackage{amsmath}
\usepackage{hyperref}
\usepackage{graphicx}
\graphicspath{{image/}}
\usepackage{amsfonts}
\usepackage{amsthm}
\usepackage{cases}
\usepackage{comment}
\usepackage{dsfont}
\usepackage{xcolor}

\theoremstyle{remark}

\hypersetup{
    colorlinks=true,       
    linkcolor=blue,          
    citecolor=magenta,        
    filecolor=red,      
    urlcolor=cyan           
}

\usepackage{color}
\definecolor{Blue}{rgb}{0.00, 0.00, 1.00}
\definecolor{Red}{rgb}{1.00, 0.00, 0.00}

\newcommand{\be}{\begin{equation}}
\newcommand{\ee}{\end{equation}}
\newcommand{\bea}{\begin{eqnarray}}
\newcommand{\eea}{\end{eqnarray}}

\def \rmd {{\mathrm{d}}}
    \newcommand{\doidoi}[2]{\href{http://dx.doi.org/#1}{#2}}

\begin{document}

\setlength{\abovedisplayskip}{5pt}
\setlength{\belowdisplayskip}{5pt}

\title{Probing the large deviations of the Kardar-Parisi-Zhang equation at short time  with an importance sampling of directed polymers in random media}

\author{Alexander K. \surname{Hartmann}}
\affiliation{Institut f{\"u}r Physik, Universit{\"a}t Oldenburg, 26111 Oldenburg, Germany}
\author{Alexandre Krajenbrink}
\affiliation{Laboratoire de Physique de l'\'Ecole Normale Sup\'erieure, PSL University, CNRS, Sorbonne Universit\'es, 24 rue Lhomond, 75231 Paris, France}
\author{Pierre Le Doussal}
\affiliation{Laboratoire de Physique de l'\'Ecole Normale Sup\'erieure, PSL University, CNRS, Sorbonne Universit\'es, 24 rue Lhomond, 75231 Paris, France}

\date{\today}

\begin{abstract}
The one-point distribution of the height for the continuum Kardar-Parisi-Zhang (KPZ) equation is determined numerically using the mapping to the directed polymer in a random potential at high temperature. Using an importance sampling approach, the distribution is obtained over a large range of values, down to a probability density as small as $10^{-1000}$  in the tails. The short time behavior is investigated and compared with recent analytical predictions for the large-deviation forms of the probability of rare fluctuations, showing a spectacular agreement with the analytical  expressions. The flat and stationary initial conditions are studied in the full space, together with the droplet initial condition in the half-space.
\end{abstract}



\maketitle

\section{Introduction}

We study the continuum Kardar-Parisi-Zhang (KPZ) equation \cite{KPZ} in 1+1 dimensions, which describes the stochastic growth of an interface parameterized by a height field $h(x,t)$
\begin{equation}
\label{eq:KPZ}
\partial_t h(x,t) = \nu \partial_x^2 h(x,t) + \frac{\lambda_0}{2} (\partial_x h(x,t))^2 + \sqrt{D} \, \xi(x,t) \;,
\end{equation}
starting from an initial condition $h(x, 0 )$. Here $\xi(x,t)$ is a centered Gaussian white noise with $\mathbb{E}[\xi(x,t)\xi(x',t')]=\delta(x-x')\delta(t-t')$, and we use from now on units of space, time and heights such that $\lambda_0=D=2$ and $\nu=1$ \cite{footnote100}. We consider here the Cole-Hopf solution, such that $Z(x,t)=e^{h(x,t)}$ satisfies the stochastic heat equation (SHE), 
\begin{equation} \label{she} 
\partial_t Z(x,t)=\partial^2_x Z(x,t) +\sqrt{2}\xi(x,t) Z(x,t)
\end{equation}
Then $Z(x,t)$ equals the partition sum of a continuum directed polymer in the $d=1+1$ random potential $\xi$,
with one endpoint at $(x,t)$. In some cases we will consider the KPZ equation in a half-space  $x\geqslant 0$, with the Neumann boundary condition $\partial_x h(x,t)|_{x=0}=A$ for all time $t$, that is
$\partial_x Z(x,t)|_{x=0}=A Z(0,t)$, where $A$ is the boundary parameter. The choice
$A=0$ corresponds to a symmetric (reflective) wall at $x=0$, while $A=+\infty$ corresponds to an infinitely repulsive wall (hard wall, or absorbing wall) imposing $Z(0,t)=0$ for all $t$.\\

Exact solutions for the probability distribution function (PDF) of the height $h(0,t)$ of the KPZ equation, valid at all time $t$, have been obtained for several initial conditions \cite{CLR10,dotsenko,SS10,ACQ11,PLDSineGordon,QuastelFlat,CLDflat,SasamotoStationary,SasamotoStationary2,BCFV}, notably \emph{flat}, \emph{droplet} and \emph{Brownian} (also called \emph{stationary}). 
They showed convergence to Tracy-Widom distributions at large time for the
typical fluctuations, depending on the class of initial condition. Exact solutions
valid at all time have also been obtained for the half-space geometry \cite{KrajenbrinkReplica,gueudre2012directed,borodin2016directed,barraquand2017stochastic}.

Recently, the large deviations away from the typical behavior have been studied. At short time, a number of results have been obtained for a variety of initial conditions. While the typical height fluctuations are
Gaussian and of scale $\delta h \sim t^{1/4}$, it was shown that the PDF of the shifted random variable $H=H(t)=h(0,t)-\langle h(0,t) \rangle$ enjoys the large-deviation principle in the regime $H = \mathcal{O}(1)$ and $t \ll 1$ 
\begin{equation} \label{ld} 
P(H,t)\sim \exp(-\frac{\Phi(H)}{\sqrt{t}})   
\end{equation}
with $\Phi(0)=0$. The rate function $\Phi(H)$ exhibits
 some universal properties : it behaves as
$\sim H^2$ for small $|H| \ll 1$, in agreement with the Gaussian form of the typical fluctuations, and has 
asymmetric tails, $|H|^{5/2}$ on the negative side and $H^{3/2}$ on the positive side, 
with amplitudes depending on the initial condition, see Table \ref{table:ShortTime_tails}.

\begin{table}[ht!]
\begin{tabular}{p{5.3cm} p{1.1 cm} p{0.3cm} p{1.2cm}}
   \hline
   Initial condition &  Left tail & & Right tail \\ [1ex]
\hline
\textbf{Full space}&    &   \\[1ex]
Droplet & $\frac{4}{15 \pi} |H|^{5/2}$  & & $\frac{4}{3} H^{3/2}$  \\[1ex]
Brownian - Stationary & $\frac{4}{15 \pi} |H|^{5/2}$  & & $\frac{2}{3} H^{3/2}$  \\[1ex]

Flat& $\frac{8}{15 \pi} |H|^{5/2}$  & & $\frac{4}{3} H^{3/2}$  \\[1ex]
\textbf{Half space}&    &&   \\[1ex]
     Droplet with reflective wall & $\frac{2}{15 \pi} |H|^{5/2}$   && $\frac{2}{3} H^{3/2}$  \\[1ex]
        Droplet with repulsive hard wall & $\frac{2}{15 \pi} |H|^{5/2}$ &  & $\frac{4}{3} H^{3/2}$  \\[1ex]
             Brownian with repulsive hard wall & $\frac{2}{15 \pi} |H|^{5/2}$ &  & $\frac{2}{3} H^{3/2}$  \\[1ex]
   \hline 
\end{tabular}
   \caption{Tails of the large-deviation function $\Phi(H)$ in \eqref{ld} for various initial conditions. In half-space, we use the shorthand notations: reflective wall for $A=0$, repulsive hard wall for $A=+\infty$.
   These results have been obtained in Refs.~\cite{le2016exact,krajenbrink2017exact,AlexKrajPHD, KrajLedou2018, krajenbrink2018large} and Refs.~\cite{Baruch,MeersonParabola,janas2016dynamical,meerson2017randomic,Meerson_Landau,Meerson_flatST,asida2019large,meerson2018large,smith2018finite,smith2019time}.}
   \label{table:ShortTime_tails}
\end{table}

There are two complementary methods to study the short-time large deviations. The first one
uses the known exact solutions to the KPZ equation and allows to obtain the rate function
$\Phi(H)$ completely analytically, see Refs.~\cite{le2016exact,krajenbrink2017exact,AlexKrajPHD, KrajLedou2018, krajenbrink2018large}.
It is thus, at this stage, restricted to the initial conditions where exact
solutions are available. The second method, the weak noise theory (also called optimal fluctuation
theory), see Refs.~\cite{Korshunov,Baruch,MeersonParabola,janas2016dynamical,meerson2017randomic,Meerson_Landau,Meerson_flatST,asida2019large,meerson2018large,smith2018finite,smith2019time}.
is more versatile but leads to differential equations which, at this stage, can only be solved numerically, 
although exact results have been obtained concerning the tails $|H| \to +\infty$, in good agreement with the first method. The first method was later extended to obtain higher order corrections in the small
time expansion in Ref.~\cite{Prolhac}.

Having determined analytically the large-deviation function $\Phi(H)$, one can ask whether it is possible to match the theoretical predictions with numerical simulations \cite{practical_guide2015}: this is an outstanding problem as large deviations are by definition extremely hard to probe. In the context of the KPZ equation, a previous work \cite{NumericsHartmann}
has investigated this question for the droplet initial condition in full space, showing a very good agreement over hundreds of decades in probability.

The aim of this paper is to extend the previous numerical effort to obtain a similar quality of agreement
with analytic predictions for other initial conditions of the KPZ equation \textit{(i)} the flat and stationary, i.e. Brownian, initial conditions in full-space and \textit{(ii)} the droplet initial condition in half space with an infinite hard wall. In addition, we aim to confirm the existence of a phase transition for the stationary initial condition. This transition,
unveiled in \cite{janas2016dynamical,Meerson_Landau,Meerson_flatST} was further analyzed and confirmed in \cite{krajenbrink2017exact} where analytic formula where obtained for the critical height $H_{c2}$  and the two branches for $\Phi(H)$ which coexist beyond the transition. Here we aim to confirm {\it both branches}, and
in particular that the "analytic branch" obtained in Ref.~\cite{krajenbrink2017exact} (which does not show any phase transition)
corresponds indeed
to the flat initial condition, as surmised in Ref.~\cite{Meerson_flatST}.\\

The outline is as follows. In Section \ref{sec:methods} we explain the methods (i) the use of a lattice directed 
polymer in the high temperature limit to simulate the continuum KPZ equation (ii) the
importance sampling method which allows to explore the deep tails of the large-deviation
regime. In Section \ref{sec:compare} we study first the flat and stationary initial conditions
(since their large deviations are related) and, secondly, the half-space KPZ with an infinitely repulsive wall.
We take the opportunity to summarize in each case the analytical predictions for the
rate function $\Phi(H)$. Next we specify how the directed polymer mapping is implemented.
Finally we show and comment the numerical results. 

\section{Methods}
\label{sec:methods}

\subsection{Directed polymer on a lattice}
\begin{figure}[ht!]
    \centering
    \includegraphics[scale=0.45]{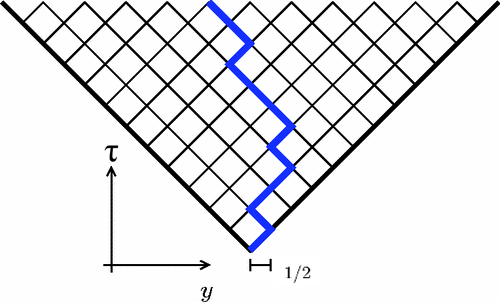}
    \caption{Representation of a directed polymer on a square lattice. The square is rotated by  $45^{\circ}$ and the path of the polymer starts from one of the corner of coordinate $(0,0)$.}
    \label{fig:lattice}
  \end{figure}

Define a directed polymer on the rotated square lattice $(y,\tau)$, see Fig.~\ref{fig:lattice}, which is allowed to grow according to the following rule 
\begin{equation}
(y,\tau)\to (y\pm \frac{1}{2},\tau+1) 
\end{equation}

  For each site of the square lattice, define a quenched
random variable $V_{y,\tau}$, a temperature $T$ and an associated Boltzmann weight $\exp(- \frac{V_{y,\tau}}{T})$. For a path $\gamma:(0,0)\to (y_f,L)$, we define its weight by
\begin{equation}
w_\gamma =\prod_{(y,\tau)\in \gamma}e^{- \frac{V_{y,\tau}}{T}}.
\end{equation}

For paths of length $L$ with a starting point $(0,0)$ and 
an end-point $(y_f,L)$ 
the partition sum over all possible polymers is given by the sum of the weight of all paths joining these points
\begin{equation}
Z_{y_f,L}=\sum_{\gamma} w_\gamma.
\end{equation}
With these definitions, the partition sum verifies a natural recursion formula which constitutes a discretized version of the Stochastic Heat Equation \cite{brunetTHESIS,reviewCorwin} 
   \begin{equation}
Z_{y,\tau+1}=\left(Z_{y-\frac{1}{2},\tau}+Z_{y+\frac{1}{2},\tau} \right) e^{-\beta V_{y,\tau+1}}   .
   \end{equation}
Numerically, this problem is solved by the \textit{transfer matrix method} and the complexity to compute the partition sum up to some discrete time $\tau$ is of order $\mathcal{O}(\tau^2)$. It is interesting to investigate the extreme regimes of zero temperature and high temperature.

At zero temperature, $T \to 0$, the free energy defined as $F_{y,\tau}=-T\log Z_{y,\tau}$ verifies the optimization/recursion equation 

\begin{equation}
F_{y,\tau+1}=\min(F_{y-\frac{1}{2},\tau}, F_{y+\frac{1}{2},\tau})+V_{y,\tau+1}\; .
\end{equation} 

In the high temperature regime, $T \to \infty$, the lattice polymer converges to the continuum one.
The continuum polymer, whose partition function is the solution of the SHE \eqref{she}, is obtained from the lattice one 
using the following parametrization \cite{CLR10,brunetTHESIS}
      \begin{equation}\label{eq:DPparamConTI}
      x=\frac{4y_f }{T^2}, \qquad t=\frac{2L}{T^4} .
      \end{equation}
The proper convergence of the partition function is expressed as
      \begin{equation}
\lim_{\substack{L\to+\infty\\ T\to +\infty\\ t \, \text{fixed}}}\hspace*{0.3cm} 2^{-L} Z_{y_f,L}= Z(x,t)
\end{equation}
If the random variables $V_{y,\tau}$ are chosen as independent unit Gaussian (which is the
choice made everywhere in this paper) then 
$Z(x,t)$ is solution of the SHE \eqref{she}.
%
      The coordinates $(x,t)$ are associated to the continuum model. From this parameterization, we see that taking the limits of infinite temperature $T$ and infinite lattice length $L$, while keeping the time $t$ fixed and small, 
      allows to access the short-time dynamics of the SHE, equivalently of the KPZ equation, 
      which is our present aim.\\

Having defined the temperature regime, we now need to choose the geometry of the polymer problem. 
We 
consider two type of configurations.
\begin{itemize}
\item The point to point polymer whose starting and ending points are fixed.
\item The point to line polymer whose starting point is fixed, but the ending point
is arbitrary on some line. In that case the partition sum is obtained by summing over all ending points on this line, with possibly an additional weight. In the limit of large temperature $T$ and the large lattice size $L$, this sum will be viewed as a Riemann sum approaching an integral.
\end{itemize}


We recall that if we want to evaluate the final KPZ height at $x=0$, then its formal expression at time $t$ is given by 
      \begin{equation}
      e^{h(0,t)}=\int_{\mathbb{R}}\mathrm{d}x' \, Z(0,t \vert \,  x',0)e^{h(x',t=0)}
      \end{equation}
where $Z(x,t \vert \,  x',0)=Z(x,t)$ denotes the solution of the SHE with 
initial condition
$Z(x,t=0  \vert\,  x',0)=\delta(x-x')$, which corresponds to the so-called droplet initial condition
(at position $x'$) for the KPZ equation. From the above, we thus need to encode the initial condition of the KPZ equation in the configuration of the polymer using the correspondence
\begin{equation}
Z(0,t \vert \,  x,0)  \longleftrightarrow 2^{-L} Z_{y_f,L}
 \end{equation} 
The usual initial conditions will be encoded as follows:
\begin{itemize}
\item The droplet initial condition is $e^{h(x,t=0)}=\delta(x)$ so $e^{h(0,t)}= Z(0,t \vert 0,0)$, this is precisely the point to point polymer;
\item The flat initial condition is $e^{h(x,t=0)}=1$ 
so $e^{h(0,t)}=\int_{\mathbb{R}}\mathrm{d}x' \, Z(0,t \vert x',0)$. This integral would be obtained on the discrete lattice by summing over all points over the initial line. It is however fully equivalent (and more convenient for our study) to flip the figure upside down, and sum over all points over the final line:
this is precisely the point to line polymer as represented in Fig~\ref{fig:lattice}, where the notion of final line is made clear (we will use the same notion for the Brownian initial condition below). 
\item The Brownian initial condition is $e^{h(x,t=0)}= e^{B(x)}$ 
so $e^{h(0,t)}=\int_{\mathbb{R}}\mathrm{d}x' \, Z(0,t \vert  x',0)
e^{B(x')}$ where $B(x')$ is the unit two-sided Brownian motion with $B(0)=0$. This integral will be obtained on the discrete lattice by summing over all points over the final line (see discussion above) with an additional weight representing the Brownian contribution, hence this is another type of point to line polymer. The exact method and scaling to introduce the Brownian contribution will be discussed later.
\end{itemize}

The analytical predictions being for the centered one-point KPZ height, we also need to center the distribution of the height obtained in the simulations. Calling $Z_\tau$ the discrete partition function up to time $\tau$ (whatever configuration), we will compare our theoretical predictions to the distribution of the quantity $H(t)$ defined by
\begin{equation}
\label{eq:Hhartmann}
H(t)=\log \frac{Z_L}{\overline{Z_L}}
\end{equation}
where $\overline{Z_L}$ is the average value of the partition function over many realizations\footnote{An additional centering to impose $\Phi(0)=0$ will be applied on the numerical data (see below). It
corresponds to corrections which are subdominant at small $t$ in the continuum model
(i.e. the difference between $\overline{\log Z(0,t)}$ and $\log \overline{Z(0,t)}$
\cite{CLR10,flatshorttime}).}.

\subsection{Introduction to importance sampling}
For the purpose of the introduction of the idea of importance sampling, we retain some elements of the presentation made in Ref.~\cite{NumericsHartmann}.  In principle one could obtain an estimate of the probability distribution
$P(H,t)$ numerically from \textit{simple
sampling}. For this, one generates many disorder realisations and calculates the partition function for each. Then  $\overline{Z_L}$ is estimated by averaging over all samples, and the
distribution is the histogram  of the values of $H$ according to Eq.~\eqref{eq:Hhartmann}. Nevertheless, this limits the smallest probabilities which can be resolved to the inverse of the number of samples, hence reaching probabilities as small as $10^{-1000}$ is strictly impossible. Hence, a different approach is required.

To estimate $P(H,t)$ for a much larger range, where probability densities as small as $10^{-1000}$ may appear, we use a more powerful approach, called  importance sampling as discussed in Refs.~\cite{align2002,largest-2011}.  This approach has been successfully applied to many problems in statistical physics and mathematics 
to obtain the tails of distributions arising in equilibrium and non-equilibrium situations \cite{rare-graphs2004,partition2005,monthus2006,rnaFreeDistr2010,driscoll2007,saito2010,fBm_MC2013,work_ising2014,convex_hull2015,convex_hull_multiple2016}. The idea behind importance sampling is to sample the different disorder realisations with
a suitable additional bias \cite{hammersley1956}. Here we use the bias $\exp(-\theta H(V))$ where $\theta$  is an adjustable parameter interpreted as a fictive inverse temperature.
Varying the value of $\theta$ allows one to sample the
distribution in different regions. In particular, if $\theta>0$ the configurations with a negative $H$ become more likely, conversely
if $\theta<0$ the configurations with a positive $H$ are favored.   
A standard Markov-chain Monte Carlo simulation is used to sample the biased configurations 
 \cite{newman1999,landau2000}. At each time step a new disorder realisation $V^*$ is proposed by replacing on the current realisation $V$ 
a certain fraction $r$ of the random numbers $V_{y,\tau}$ by new random numbers.
The new disorder realisation is then accepted with the usual Metropolis-Hastings probability
\begin{equation}
p_{\rm Met} = \min\lbrace1,e^{-\theta\left[H(V^*)-H(V)\right]}\rbrace,
\end{equation}
otherwise the old configuration is kept \cite{metropolis1953}.
Note that the average partition function $\overline{Z_L}$ appearing
in the definition of $H$ \eqref{eq:Hhartmann} drops out of the Metropolis probability.
 By construction, the algorithm fulfils detailed balance and is ergodic, since within a sufficient number of steps, each possible realisation may be constructed. Thus,
in the limit of infinitely long Markov chains, the distribution of biased disorder realisations will follow the probability
\begin{equation}
q_\theta(V) = \frac{1}{Q(\theta)} P_{\text{dis}}(V)e^{-\theta H(V)}\,, \label{eq:qT}
\end{equation}
where $ P_{\text{dis}}(V)$ is the original disorder distribution and $Q(\theta)= \sum_V P_{\text{dis}}(V)e^{-\theta H(V)} $ is the normalization factor. Note that $Q(\theta)$ also depends
on $L$ and $T$ because of finite size and temperature effects. $Q(\theta)$ is generally unknown but can be determined 
from the numerical results, see Appendix \ref{app:sampling}. The output of this Markov chain allows to construct a biased histogram $ P_\theta(H,t)$. In order to get the correct empirical probability density $P(H,t)$ one should debias the result so that
\begin{equation}
 P(H,t) =  e^{\theta H} Q(\theta) P_\theta(H,t).
\label{eq:rescaling}
\end{equation}
Hence, the target distribution $P(H,t)$ can be estimated, up to a normalisation
constant $Q(\theta)$. For each value of the
parameter $\theta$, 
a specific range of the distribution $P(H,t)$ will be sampled and  using a positive (respectively, negative) parameter allows to sample the region of a distribution at the left (respectively, at the right) of its center.

\section{Comparison of the theoretical predictions with the simulations}
\label{sec:compare} 

We now compare the theoretical predictions for the probability distribution of the solutions to the Kardar-Parisi-Zhang equation with the numerical simulations of the directed polymer on a lattice for each initial condition. 
We start with the stationary (i.e. Brownian) and flat initial conditions, since their analytical expressions
are related. Then we address the half-space geometry. We proceed as follows. In each case
we first recall and give explicitly the analytical predictions for the large-deviation rate function $\Phi(H)$.
Then we provide details on how to encode the random weights on the lattice model to account for the particular initial conditions. Finally we present the numerical data and compare with the predictions. The large-deviation function $\Phi(H)$ has been constructed so that it is centered around 0 and hence for a fair comparison, the probability densities obtained in the simulations will be shifted so that their maximum is also reached at 0. Finally, we insist on the fact that the comparison will be done without any fitting parameter.

\subsection{The flat and stationary initial conditions in full-space}

\subsubsection{Recall of the analytical result for $\Phi(H)$}

We first recall the results of Ref. \cite{krajenbrink2017exact} (see also Ref.~\cite{AlexKrajPHD})
for the two branches of $\Phi(H)$. As discussed below, they provide the short-time probability distribution for the flat and the stationary (i.e. Brownian) initial condition in full-space.
Let us first define the critical height $H_{c2}=2\log(2e -\mathcal{I})-1\simeq 1.85316$ with
\begin{equation}
\mathcal{I}=\int_0^{+\infty}\frac{\mathrm{d}y}{\pi}\left[1+\frac{1}{y}\right]\frac{\sqrt{y}}{e^{-1}+ye^y}.
\end{equation}
together with the function $\Psi(z)$ expressed as
\begin{equation}
\Psi(z)= -\int_\mathbb{R} \frac{\rmd k }{2\pi}{\rm Li}_2(-\frac{ze^{-k^2}}{k^2}),
\end{equation}
and its reduced version $\psi(z)=\Psi(z)-2z\Psi'(z)$. We further denote the intervals:
\begin{equation}
\begin{split}
I_1&= \left[0,+\infty\right[, \; I_2= \left[0,e^{-1}\right], \; I_3= \left]0,e^{-1}\right],\\
J_1&=\left]-\infty,0\right], \; J_2=\left[0,H_{c2}\right], \; J_3=\left[H_{c2},+\infty\right[.
\end{split}
\end{equation}
We define a large-deviation function $\Phi(H)$ given in a parametric form by 
\begin{equation}\label{eq:Brownian123456}
\begin{split}
e^H=&
\begin{cases}
z\Psi'(z)^2, &z\in I_1, H\in J_1\\
 z\left[\Psi'(z)-2z^{-1}[-W_0(-z)]^{\frac{1}{2}}\right]^2\hspace*{-0.2em}, \hspace*{-0.2cm} &  z\in I_2, H\in J_2
\end{cases}\\
\Phi(H)=&
\begin{cases}
\psi(z), &z\in I_1\\
\psi(z)+\frac{4}{3}[-W_0(-z)]^{\frac{3}{2}},  \hspace*{1.25cm}& z\in I_2
\end{cases}
\end{split}
\end{equation}
For $z\in I_3, H\in J_3$, we define two continuations of $\Phi(H)$: a symmetric analytic one (resp. asymmetric non-analytic)  denoted $\Phi_{\bf A}$ (resp.~$\Phi_{\bf NA}$) such that their parametric representations read
\begin{equation}\label{eq:Brownian1234}
\begin{split}
&\begin{cases}
e^H=z\left[\Psi'(z)-2z^{-1}[-W_{-1}(-z)]^{\frac{1}{2}}\right]^2\\
\Phi_{\bf A}(H)=\psi(z)+\frac{4}{3}[-W_{-1}(-z)]^{\frac{3}{2}}   
\end{cases}\\
&\hspace*{0.2cm} \text{and}\\
&\begin{cases}
e^H=z\left[\Psi'(z)-z^{-1}\big([-W_{-1}(-z)]^{\frac{1}{2}}+[-W_{0}(-z)]^{\frac{1}{2}}\big)\right]^2\\
\Phi_{\bf NA}(H)=\psi(z)+\frac{2}{3}[-W_{-1}(-z)]^{\frac{3}{2}}   +\frac{2}{3}[-W_{0}(-z)]^{\frac{3}{2}}   
\end{cases}
\end{split}
\end{equation}
The large-deviation solutions for the flat and the Brownian initial conditions are by far the most difficult to study, as two successive continuations are required (while only one is required for the droplet initial condition
\cite{le2016exact}). Indeed, the large-deviation functions for the stationary and flat initial conditions in full-space are given for all $H$ in $\mathbb{R}$ by
\begin{equation}\label{eq:ShortTimeFlatBrownian}
 \Phi_{\rm Brownian}(H)= 
 \Phi_{\bf NA}(H)
 , \quad \Phi_{\rm flat}(H)=2^{-\frac{3}{2}}\Phi_{\bf A}(2H).
\end{equation}
The indices $\bf A$ and $\bf NA$ are explicitly indicated for the region $H\in \left[H_{c2},+\infty\right[$ where the two continuations are distinct and can be omitted for the region where $H\in \left]-\infty,H_{c2}\right]$ where the two functions are the same, i.e. $\Phi_{\bf NA}(H)=\Phi_{\bf A}(H)$ for $H\in \left]-\infty,H_{c2}\right]$.
We further recall that $W_0$ and $W_{-1}$ are the two real branches of the Lambert function, see Appendix~\ref{app:lambert}. 

The result \eqref{eq:ShortTimeFlatBrownian} for the Brownian initial condition which follows the ${\bf NA}$ branch was obtained in Ref.~\cite{krajenbrink2017exact}. The result \eqref{eq:ShortTimeFlatBrownian} for the flat initial condition arises from (i) the identification made in the context of the WNT in Ref.~\cite{Meerson_flatST} of the analytic branch and the large-deviation function of the flat initial condition (the rationale being that no phase transition is expected for the flat initial initial condition)
and (ii) the analytic result for the ${\bf A}$ branch in Eq.~\eqref{eq:Brownian1234} 
obtained in Ref.~\cite{krajenbrink2017exact}. The verification of \eqref{eq:ShortTimeFlatBrownian} will be an important point to seek in numerical simulations. Before we introduce the results of the simulations obtained through the importance sampling method, we will explain in some details the lattice construction of these initial conditions.
 
\begin{figure*}[ht!]
\centering
\includegraphics[trim={0 5cm 0 5cm},clip,width=0.5\linewidth]{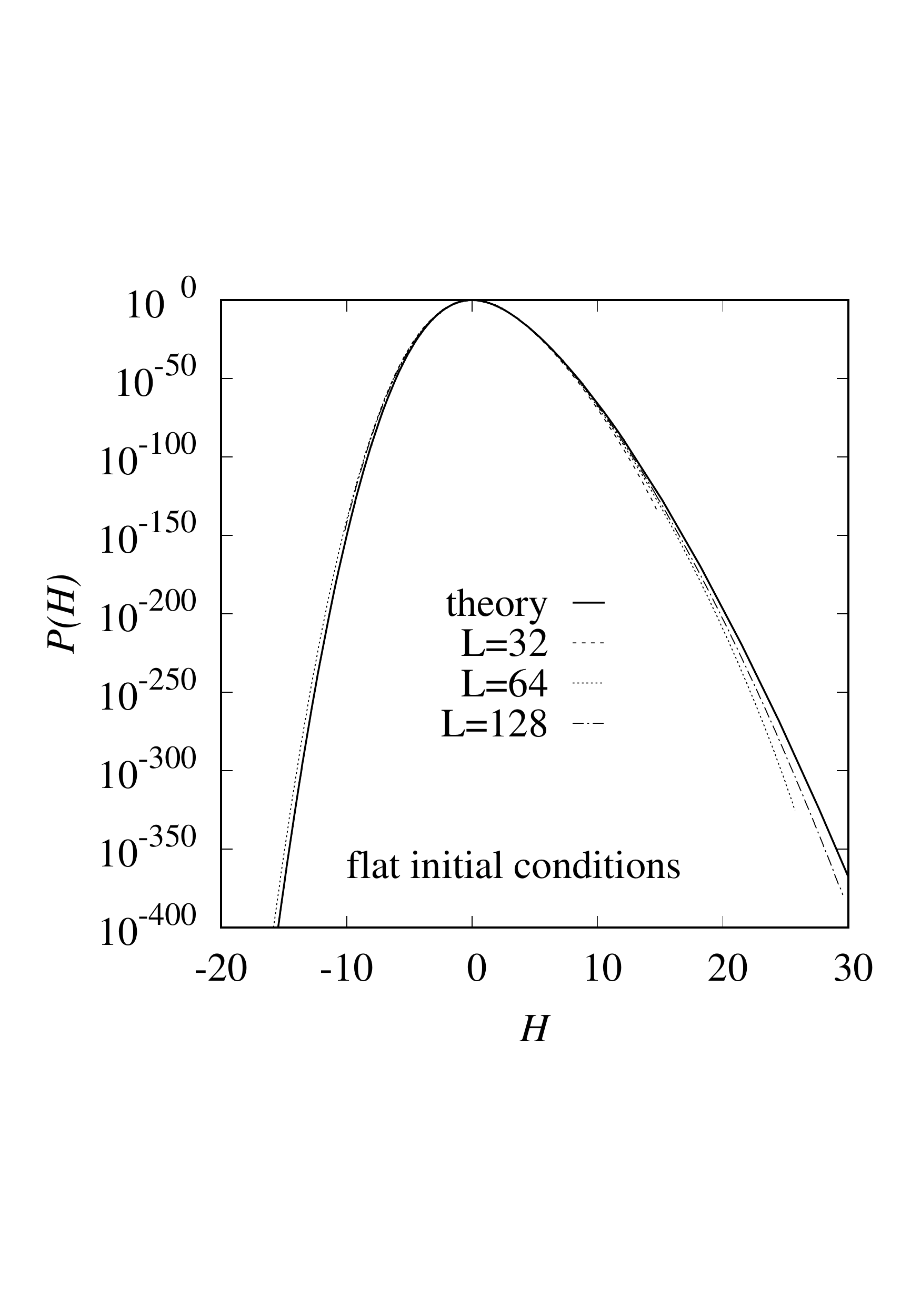}~
\includegraphics[trim={0 5cm 0 5cm},clip,width=0.5\linewidth]{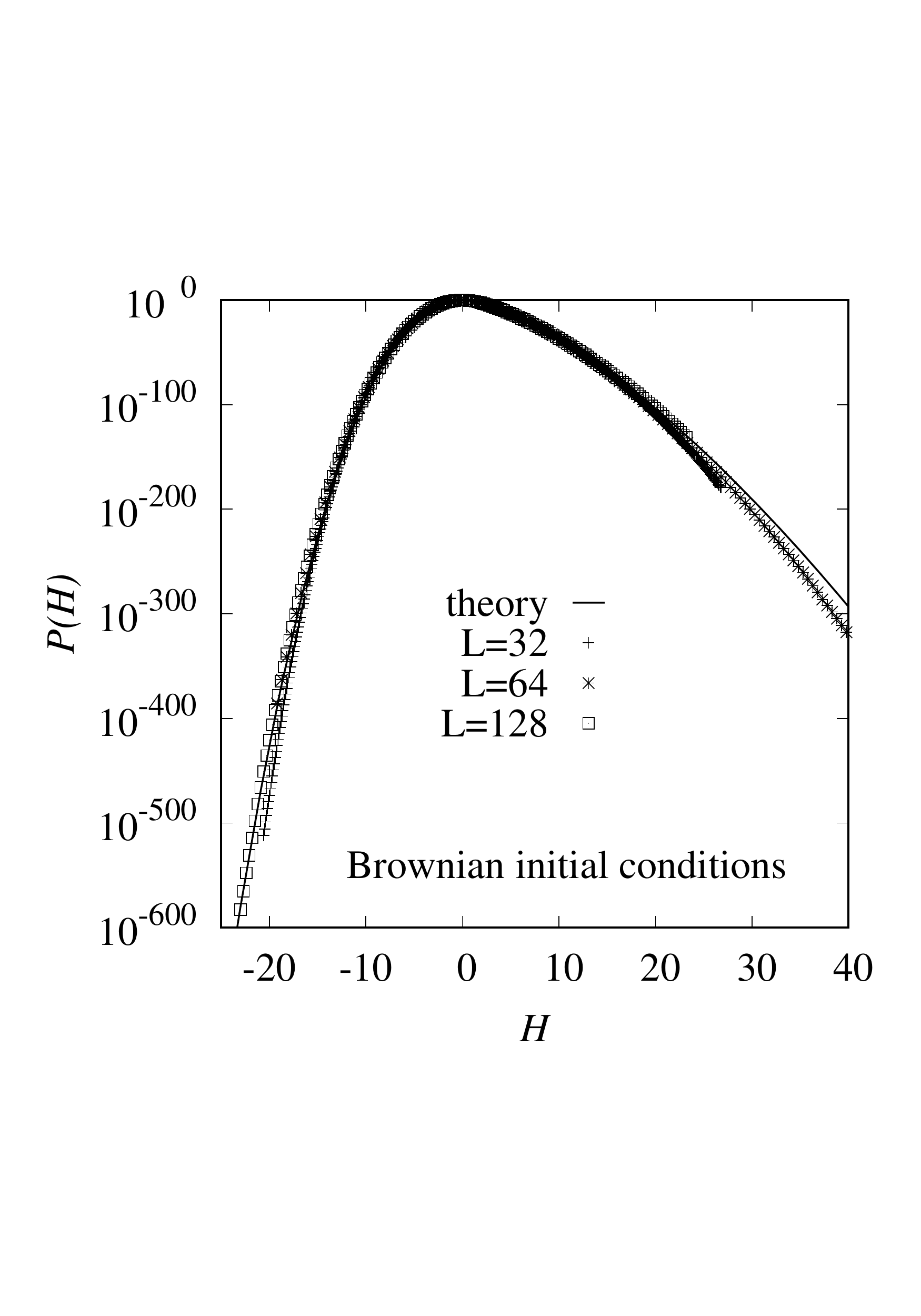}
\caption{Probability distribution $P(H,t)$ for a short time $t=1/16$ for three
different polymer lengths $L=\lbrace 32,64,128 \rbrace$ and {\bf left:} flat initial conditions,
{\bf right:} Brownian (stationary) initial conditions. In both plots the solid lines
indicate the analytical prediction displayed in Eq.~\eqref{eq:ShortTimeFlatBrownian},
together with Eqs. \eqref{eq:Brownian123456} and \eqref{eq:Brownian1234}.
\label{fig:bulkFlatBrownian}}
\end{figure*}  
 
\subsubsection{Point to line directed polymer mapping}
We first introduce 
the mapping of the flat initial condition of the KPZ equation to the directed polymer model on the lattice and then extend the discussion to the Brownian initial condition. As mentioned earlier, contrary to the droplet initial condition, the flat initial condition requires to perform a summation of the partition sum over the final line of the lattice, \textit{i.e.}
\begin{equation}
Z_L=\sum_{k=-L}^L Z_{y_k,L}.
\end{equation}
where the points $\lbrace y_k \rbrace$ are all the points over the final line of the lattice ordered from left to right. Note that this amounts to set a uniform weight over the final line. Comparatively, for the droplet initial condition, all the weight was concentrated at $k=0$ (the above sum with a Kronecker delta at $k=0$ would yield the droplet initial condition).

Recalling the continuum parametrization of Eq.~\eqref{eq:DPparamConTI}, the infinite temperature limit (up to a proportionality constant)
\begin{equation}
Z_{y_k,L}\longleftrightarrow Z(x=\frac{4y_k}{T^2},t=\frac{2L}{T^4})\; ,
\end{equation}
by a Riemann summation using a step of size $\Delta x=\frac{4}{T^2}\ll1$ we obtain the partition function of the flat condition (up to a proportionality constant)
\begin{equation}
\frac{4}{T^2}\sum_{k=-L}^L Z_{y_k,L}\underset{T\gg1}{\longrightarrow}\int_{-4L/T^2}^{4L/T^2} \mathrm{d}x\, Z(x,t) \; .
\end{equation}

We emphasize that because of our construction of the height \eqref{eq:Hhartmann}, the proportionality constants are irrelevant. As we renormalize the partition sum by its average according to Eq.~\eqref{eq:Hhartmann}, the prefactor $4/T^2$ can be discarded and will play strictly no role in the numerics. To approximate the integral as being over the full real line, the factor $4L/T^2$ should be taken as large as possible: its finiteness might induce additional finite size and discretization effects as compared to the point to point continuum polymer. To provide orders of magnitude, some of the simulations 
will have a time $t=1/16$ and the factor $4L/T^2$ is then in the range
 $[4,8]$ for the different lattice sizes used.

The extension of this mapping to the stationary initial condition is obtained by adding Brownian weights $e^{B(x)}$ on the final line on the lattice
\begin{equation}
Z_{y_k,L}e^{B( \frac{4y_k}{T^2})}\to Z(x=\frac{4y_k}{T^2},t)e^{B(x)} \; .
\end{equation}
By self-similarity, we have $B(\frac{4y_k}{T^2})\equiv \frac{2}{T}B(y_k)$ and hence $B(y_k)$ can be easily sampled by a random walk with Gaussian increments with unit variance. Note that the final partition function assumes an additional average over the Brownian motion compared to a deterministic initial condition, hence this provides an extra challenge on the numerical side.

 \begin{figure*}[t!]
\centering
\includegraphics[trim={0 4.5cm 0.2cm 5cm},clip,width = 0.4\linewidth]{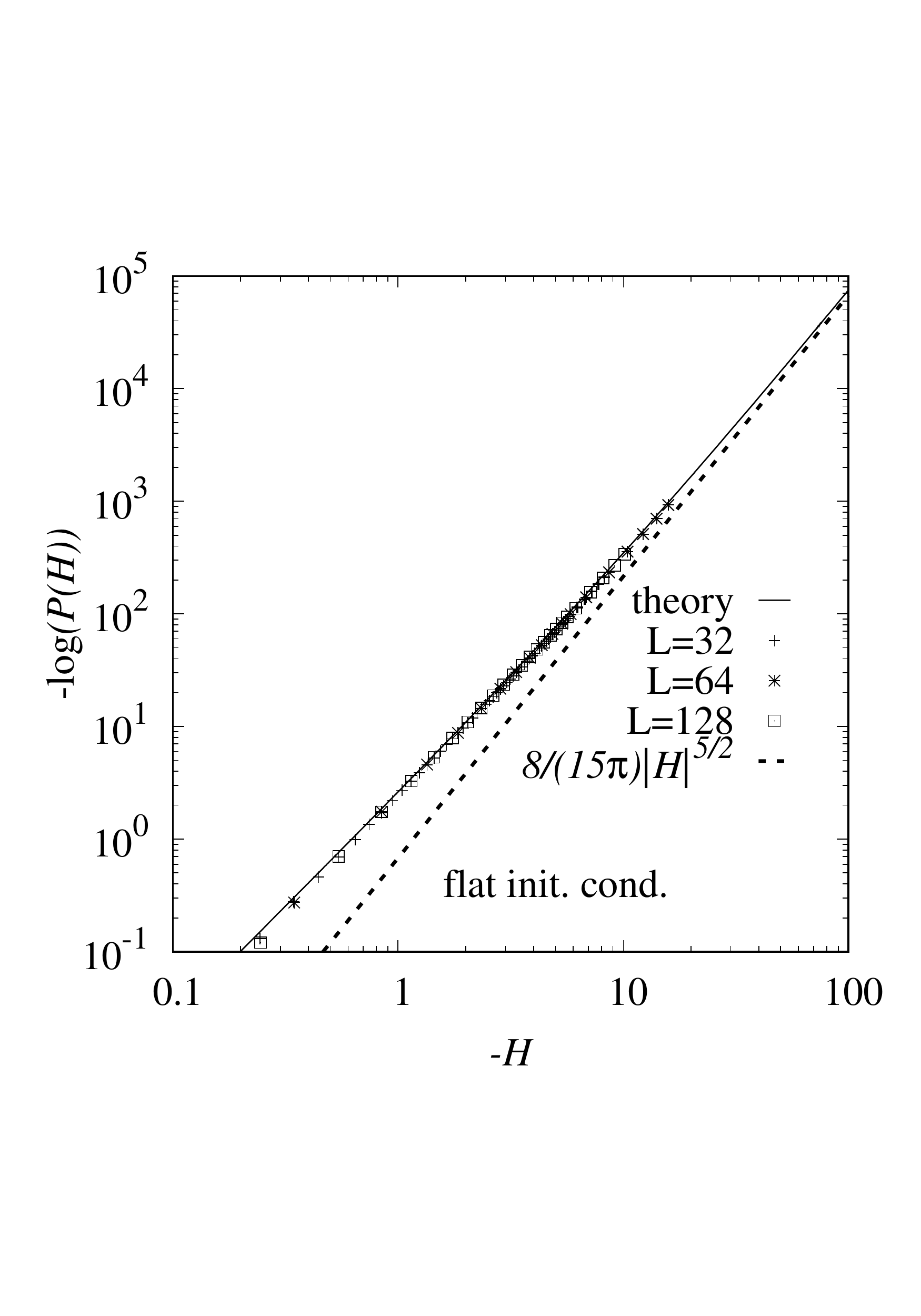}~
\includegraphics[trim={0.2cm 4.5cm 0 5cm},clip,width = 0.4\linewidth]{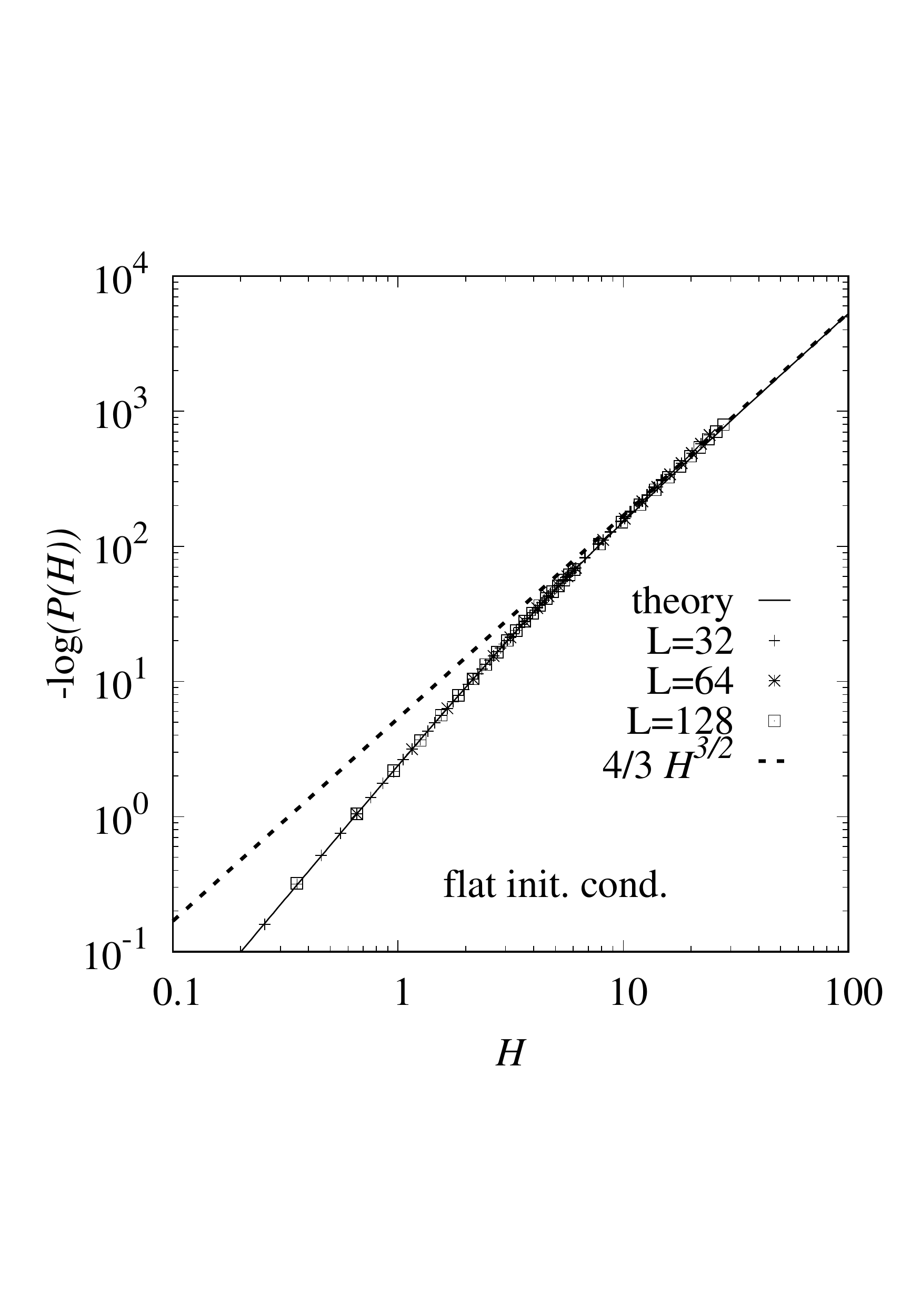}
\includegraphics[trim={0 4.5cm 0.2cm 5cm},clip,width = 0.4\linewidth]{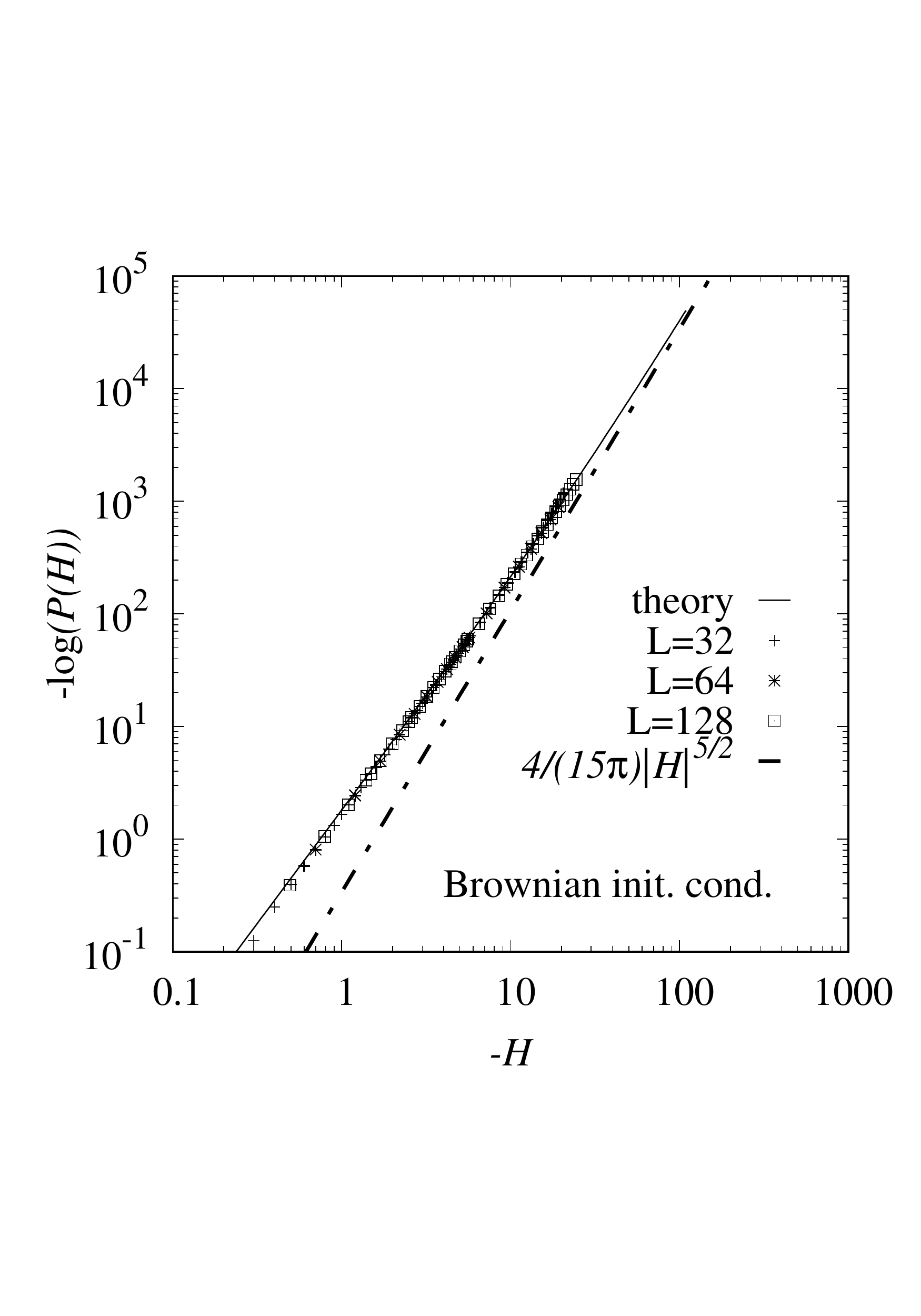}~
\includegraphics[trim={0.2cm 4.5cm 0 5cm},clip,width = 0.4\linewidth]{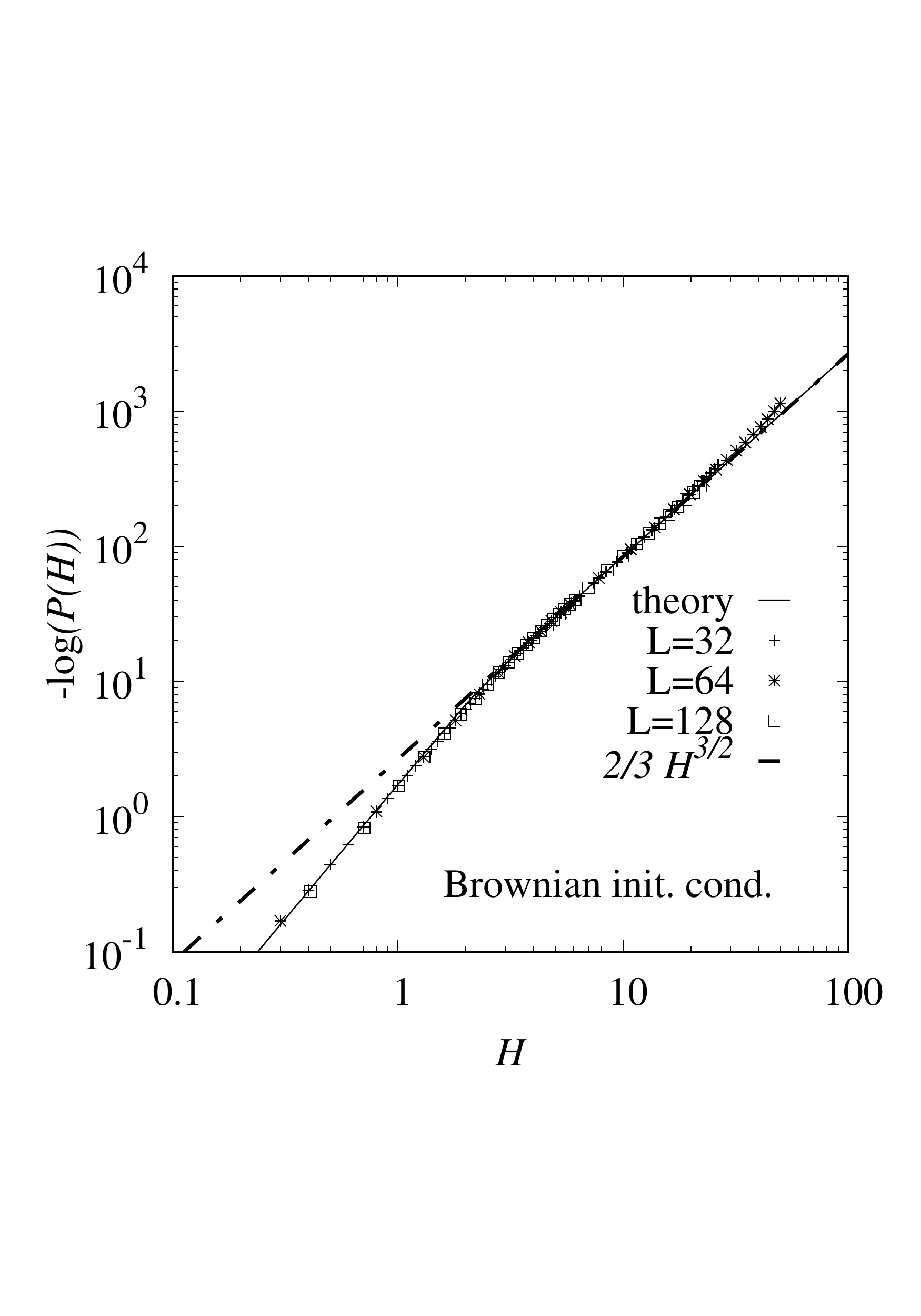}
\caption{ {\bf Top}: blow up of the left and right tails of the data shown in the left plot of Fig.~\ref{fig:bulkFlatBrownian} for the flat initial condition. {\bf Bottom}: blow up of the left and right tails of the data shown in the right plot of  Fig.~\ref{fig:bulkFlatBrownian} for the Brownian initial condition. In all four plots the solid line is the analytical
prediction given in Eq.~\eqref{eq:ShortTimeFlatBrownian},
together with Eqs. \eqref{eq:Brownian123456} and \eqref{eq:Brownian1234}. The leading tails at large $|H|$ given in the Table~\ref{table:ShortTime_tails} are also plotted in each case.}
\label{fig:tailFlatBrownian}
\end{figure*}
\subsubsection{Presentation of the simulations}

 The numerical simulations for the flat and Brownian initial conditions in full-space were run for polymers of length $L=\lbrace  32,64,128 \rbrace$ and temperature $T$ chosen so that the corresponding time for the Kardar-Parisi-Zhang equation is fixed at $t=1/16$. Convergence to the analytic predictions is expected for $L\to +\infty$. We present the simulations in Figs.~\ref{fig:bulkFlatBrownian} and \ref{fig:tailFlatBrownian}.  We observe for the flat and Brownian initial conditions that the agreement between the numerics and the theory is fairly good and improves as $L$ increases. We should note that there are additional sources of finite size effects and statistical errors since for the flat and the Brownian initial conditions we need to perform a summation of the partition function over the final line in both cases and an additional  average over the Brownian weights in the second case. Nonetheless, the simulations are able to probe both tails of the distribution of $P(H,t)$ for quite a range of values for both initial conditions as can be seen in Fig.~\ref{fig:tailFlatBrownian}. A very
good agreement of analytical and numerical results is visible here and
on this scale only very small finite-size corrections are visible. 
The numerical results even extend to the regime below $P(H)\sim e^{-1000}$ 
where the leading tails behavior starts to be dominant.


\subsubsection{Discussion around the choice of branch $\Phi_{\bf A}/\Phi_{\bf AN}$ for flat and Brownian initial conditions}
We now turn to the crucial discussion of the choice of branch $\Phi_{\bf A}/\Phi_{\bf AN}$ for both flat and Brownian initial conditions.  We recall that considerations coming from Weak Noise Theory ~\cite{Meerson_flatST}
together with the precise study of the Fredholm determinant associated to the solution for the Brownian initial condition 
\cite{krajenbrink2017exact}, concluded that for all $H$ in $\mathbb{R}$ the 
prediction \eqref{eq:ShortTimeFlatBrownian} holds, where the two branche coincide $\Phi_{\bf NA}(H)=\Phi_{\bf A}(H)$ for
$H < H_{c2}$, but differ for $H > H_{c2}$. 
This implies that if we rescale the simulation data for the flat initial condition, they should coincide pointwise up to the critical height $H_{c2}$ and then separate to the symmetric and asymmetric branches. We emphasize that the choice of branch is critical as it determines the tails of the distribution and the existence of a singularity in the large-deviation function.

\begin{figure*}[ht!]
    \centering
    \includegraphics[scale=0.725]{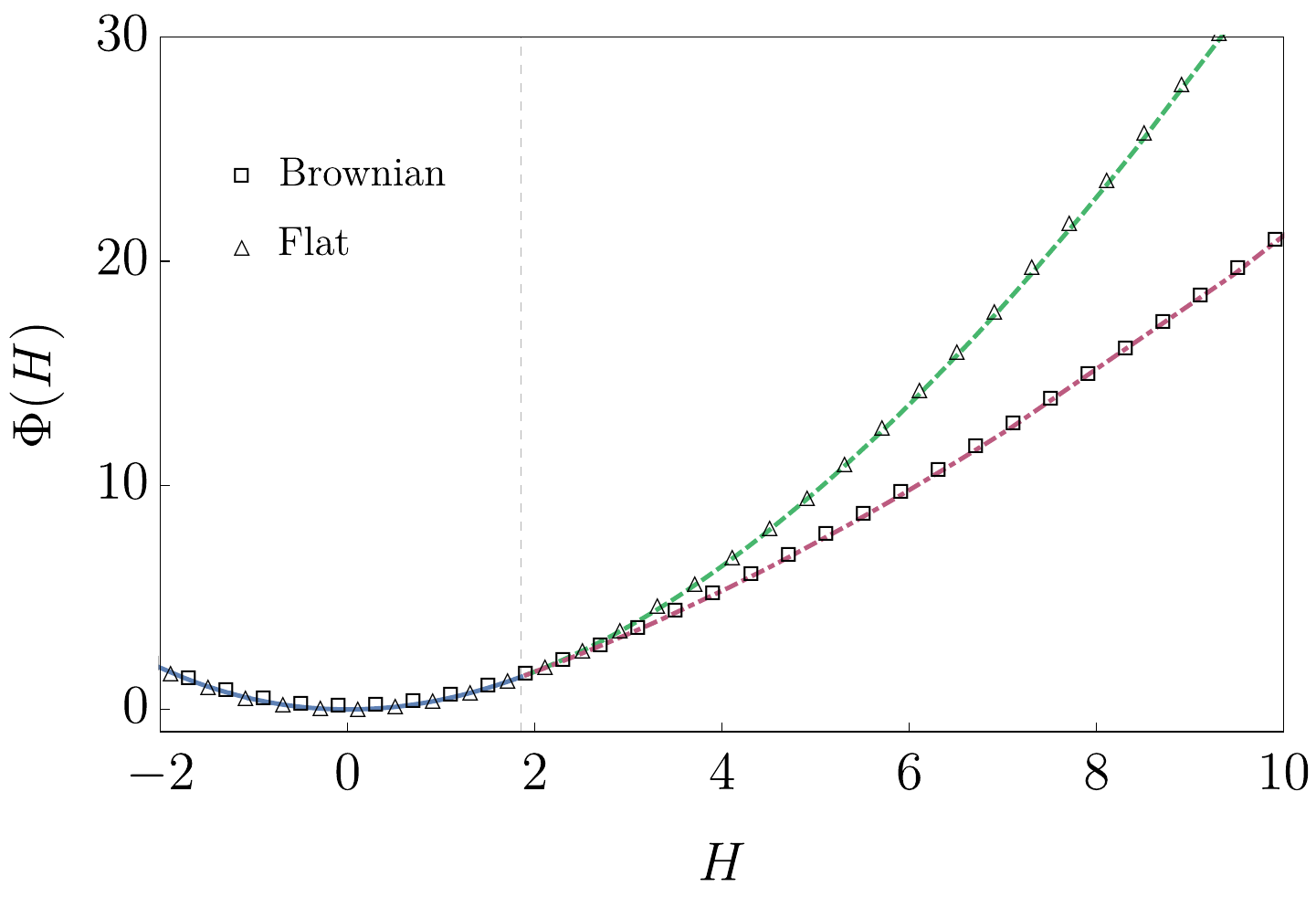}
    \caption{(color online) The analytic expression of $\Phi(H)$ obtained in Eqs.~\eqref{eq:Brownian123456} and \eqref{eq:Brownian1234} is compared with the simulation data obtained for the flat and for Brownian initial conditions. In the case of the flat initial condition, the data are rescaled according to Eq.~\eqref{eq:ShortTimeFlatBrownian} for a fair comparison. The blue line corresponds to the function $\Phi(H)$ before the critical point $H_{c2}\simeq 1.85316$ which position is indicated by a vertical grey dotted line. The green dashed line corresponds to the symmetric analytic branch $\Phi_{\bf A}$ and the red dot-dashed line corresponds to the asymmetric non-analytic branch $\Phi_{\bf NA}$. The square markers represent the Brownian initial condition simulation data and the triangle markers represent the rescaled flat initial condition data.}
          \label{first_comparison}
  \end{figure*}
  
We present in Fig.~\ref{first_comparison} the comparison in a 
region around $H_{c2}$ between the exact expression of the different branches of $\Phi(H)$ given in Eqs.~\eqref{eq:Brownian123456} and \eqref{eq:Brownian1234} and the directed polymer simulations of length $L=128$ at a time $t=1/16$ for the Brownian and the flat initial conditions. Note that we rescaled the distribution for the flat initial condition according to the scaling predicted in Eq.~\eqref{eq:ShortTimeFlatBrownian} for a fair comparison. Figure~\ref{first_comparison} confirms in a remarkable way that each initial condition corresponds to the one of the two branches that we have obtained analytically. The coincidence of the two distributions before the point $H=H_{c2}$ is also quite spectacular.

\subsection{The droplet initial condition in half-space with a hard-wall $A=+\infty$}

We now consider the KPZ equation in the half-space $x \geqslant 0$ with the boundary condition
$\partial_x h(x,0)|_{x=0}=A$, equivalently $\partial_x Z(x,t)|_{x=0}=A Z(0,t)$ for the continuum polymer, with droplet initial conditions i.e. $Z(x,t=0)=\delta(x-\epsilon)$ with $\epsilon \to 0^+$.  The large-deviation function $\Phi(H)$ was obtained in 
\cite{krajenbrink2018large}  for several values of $A$ (respectively $-1/2$, $0$ and $+\infty$).
Here we will test only the analytical prediction of Ref. \cite{krajenbrink2018large} for $A=+\infty$, which corresponds to the infinitely repulsive hard wall (i.e.
$Z(0,t)=0$ for all $t$).

%
%
%
%
%

\subsubsection{Recall of the analytical result for $\Phi(H)$}

To provide the short-time probability distribution for the droplet initial condition in a half-space in presence of a hard wall, let us first define the branching height  $H_{c}=\log(8\sqrt{\pi}\, \mathcal{I})\simeq 0.9795$ with
\begin{equation}
\mathcal{I}=\int_0^{+\infty}\frac{\mathrm{d}y}{\pi}\left[1-\frac{1}{y}\right]\frac{\sqrt{y}}{e^y/y-e}.
\end{equation}
We also define the functions $\Psi(z)$, $\Delta(z)$ and its derivative $\Delta'(z)$ as

\begin{equation}
\begin{split}
\Psi(z)&=-\int_\mathbb{R} \frac{\rmd k}{4\pi} {\rm Li}_2(-zk^2 e^{-k^2})\\
\Delta(z)&=\frac{2}{3}\left[ W_{0}(\frac{1}{z})\right]^{\frac{3}{2}}+\frac{2}{3}\left[ W_{-1}(\frac{1}{z})\right]^{\frac{3}{2}}\\
& \hspace*{1.4cm}+2\left[ W_{0}(\frac{1}{z})\right]^{\frac{1}{2}} +2\left[ W_{-1}(\frac{1}{z})\right]^{\frac{1}{2}}\\
\Delta'(z)&=-\frac{1}{z}\left[W_0(\frac{1}{z})\right]^{\frac{1}{2}}-\frac{1}{z}\left[W_{-1}(\frac{1}{z})\right]^{\frac{1}{2}}
\end{split}
\end{equation}
and we define their reduced versions $\psi$ and $\delta$
\begin{equation}
\psi(z)=\Psi(z)-z\Psi'(z), \quad \delta(z)=\Delta(z)-z\Delta'(z).
\end{equation}
We further denote the intervals:
\begin{equation}
\begin{split}
I_1&= \left[-e,+\infty\right[, \quad I_2= \left[-e,0\right[,\\
J_1&=\left]-\infty,H_c\right], \quad J_2=\left[H_c,+\infty\right[.
\end{split}
\end{equation}
The associated large-deviation function $\Phi(H)$ is expressed by the parametric system

\begin{equation}\label{eq:STdropletHWHW}
\begin{split}
e^H=&
\begin{cases}
8\sqrt{\pi}\Psi'(z), &z\in I_1, H\in J_1\\
8\sqrt{\pi}[\Psi'(z)+\Delta'(z)], & z\in I_2, H\in J_2
\end{cases}\\
\Phi(H)=&
\begin{cases}
\psi(z), &z\in I_1\\
\psi(z)+\delta(z),\hspace*{1.25cm}  & z\in I_2
\end{cases}
\end{split}
\end{equation}

Before we introduce the results of the simulations obtained through the importance sampling method, 
we will explain in some details the lattice construction of the half-space problem in the presence of a wall.

\subsubsection{Point to point directed polymer mapping in a half-space with a hard-wall $A=+\infty$}
  \begin{figure}[ht!]
    \centering
    \includegraphics[scale=0.45]{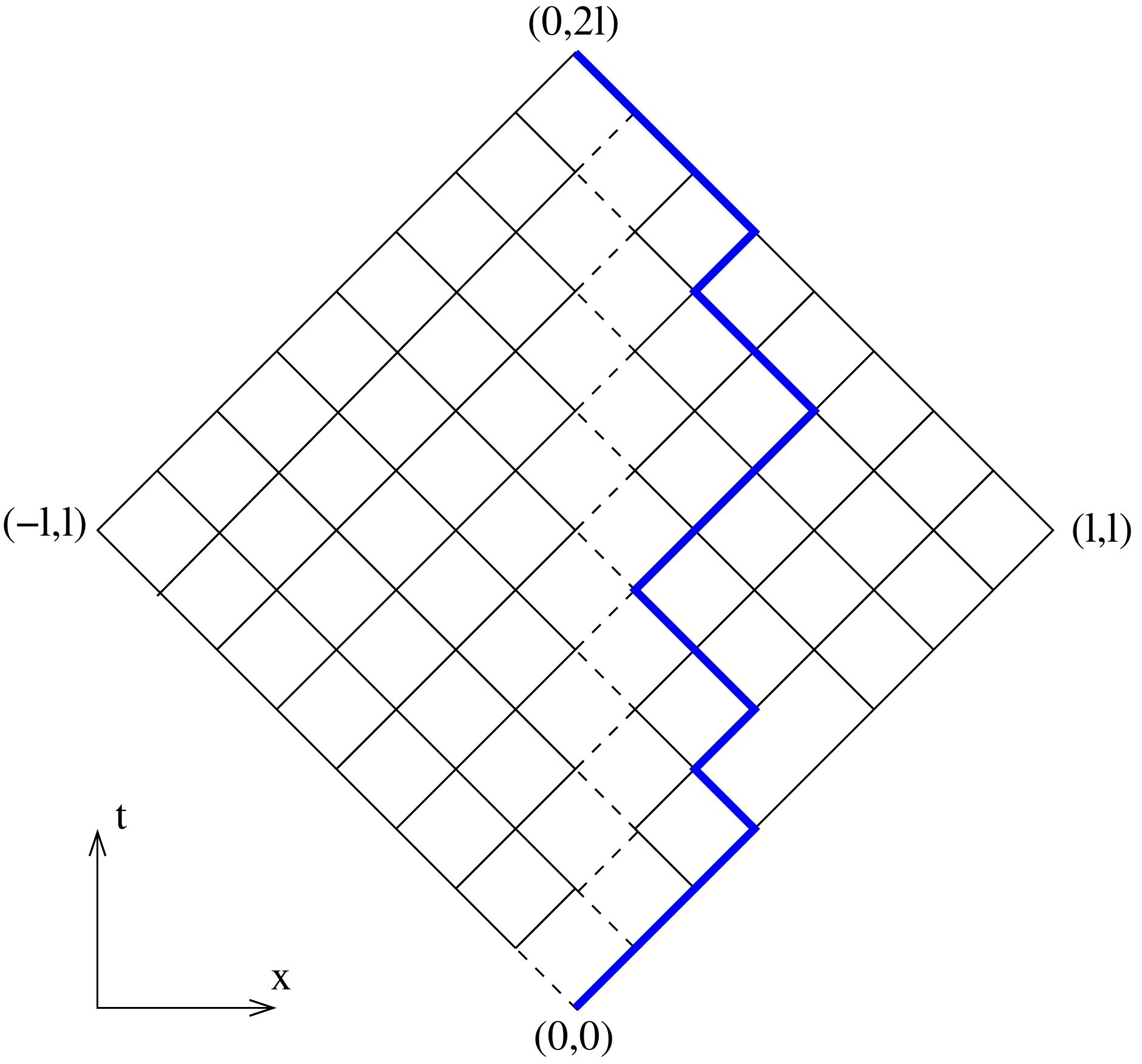}
    \caption{Square lattice designed for the half-space problem with a hard-wall. The dashed lines are forbidden edges for the polymer constraining it to stay on the right of the lattice. An example of a polymer realization is drawn on the blue line. In this representation, we have $L=2\ell$.}
    \label{fig:lattice_hardwall}
  \end{figure}
To mimic a hard-wall type problem, one can forbid the polymer to visit some edges and introduce a tabu list for edges bringing the polymer towards the diagonal of the lattice. Indeed, the partition function should be strictly zero on the diagonal so that the probability of crossing the diagonal is strictly zero. We introduce this idea of forbidden edges in Fig.~\ref{fig:lattice_hardwall}. Finally, let us mention that an extension of this construction to a generic value of $A$ would be interesting.

\subsubsection{Presentation of the simulations}
 \begin{figure*}[ht!]
\centering
\includegraphics[trim={0 4.5cm 0 5cm},clip,width = 0.38\linewidth]{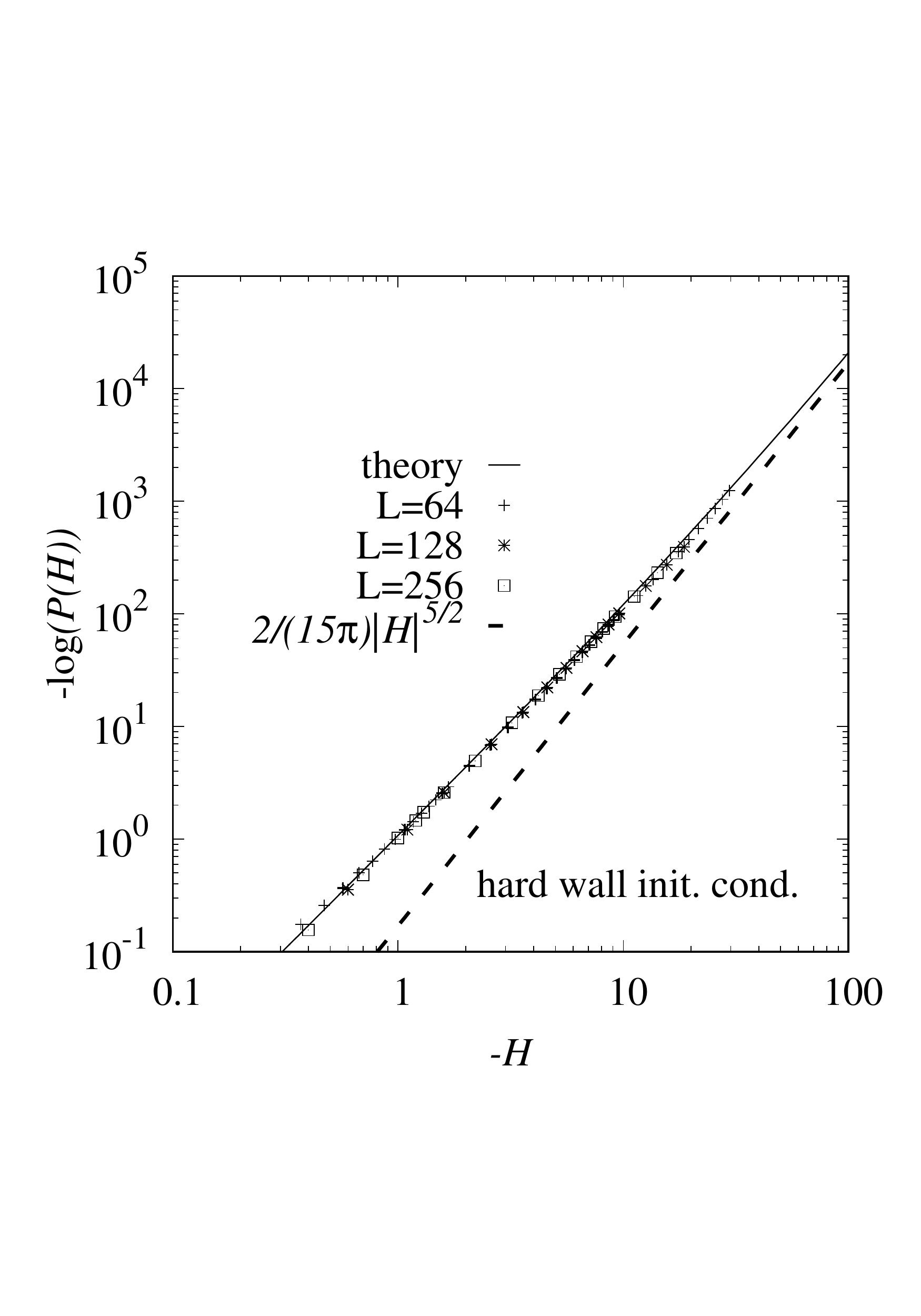}~
\includegraphics[trim={0 4.5cm 0 5cm},clip,width = 0.38\linewidth]{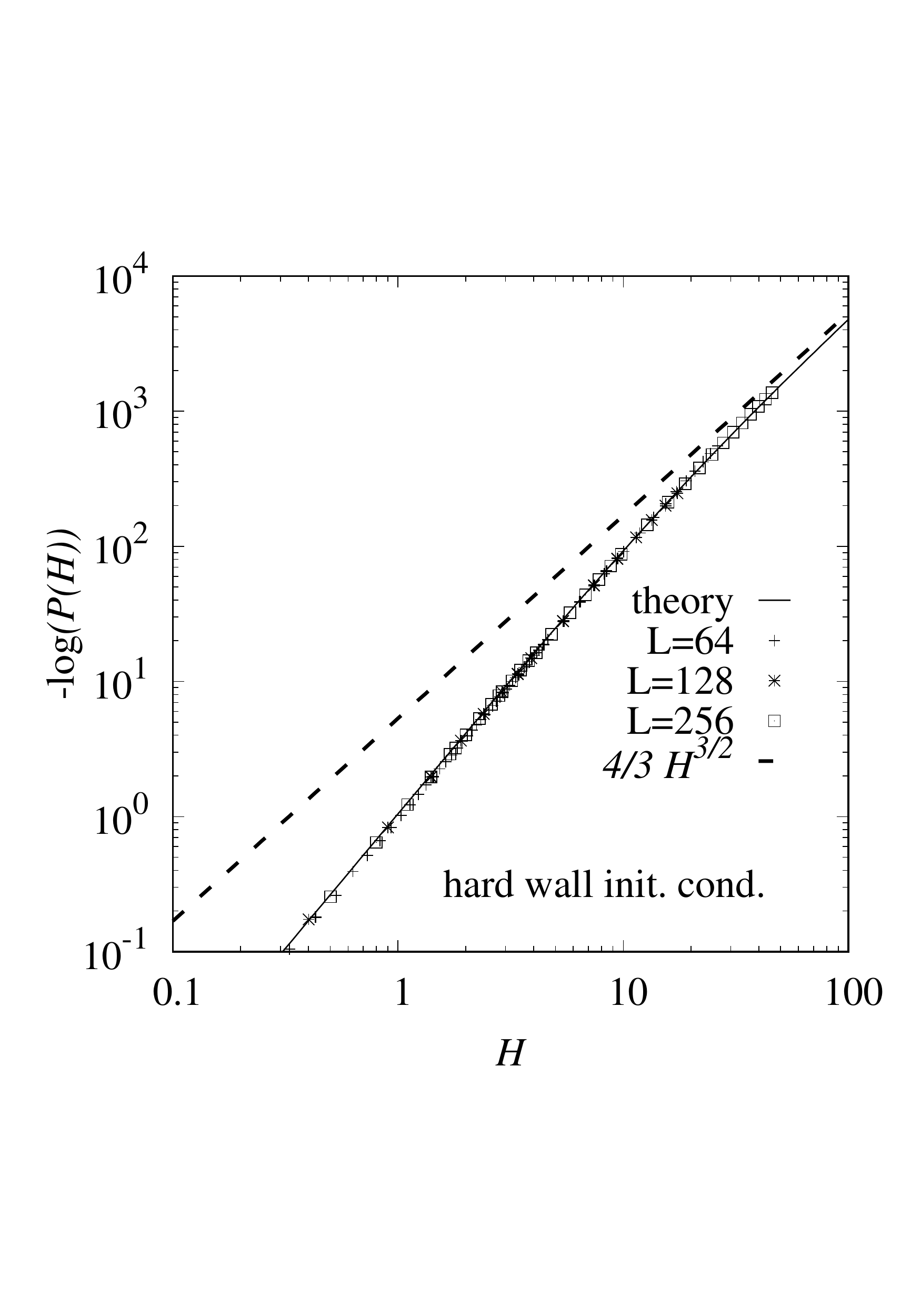}
\caption{  Blow up of the left and right tails of the data shown of Fig.~\ref{fig:P_H_0625HW1}.
The solid lines indicate the analytical prediction displayed in Eq. \eqref{eq:STdropletHWHW} obtained 
in \cite{krajenbrink2018large}. The data is also compared with the leading behavior of each
tail as displayed in the Table~\ref{table:ShortTime_tails}.
\label{fig:P_H_0625HW2}}
\end{figure*}

\begin{figure}[ht!]
\centering
\includegraphics[trim={0 5cm 0 5cm},clip,width=0.95\linewidth]{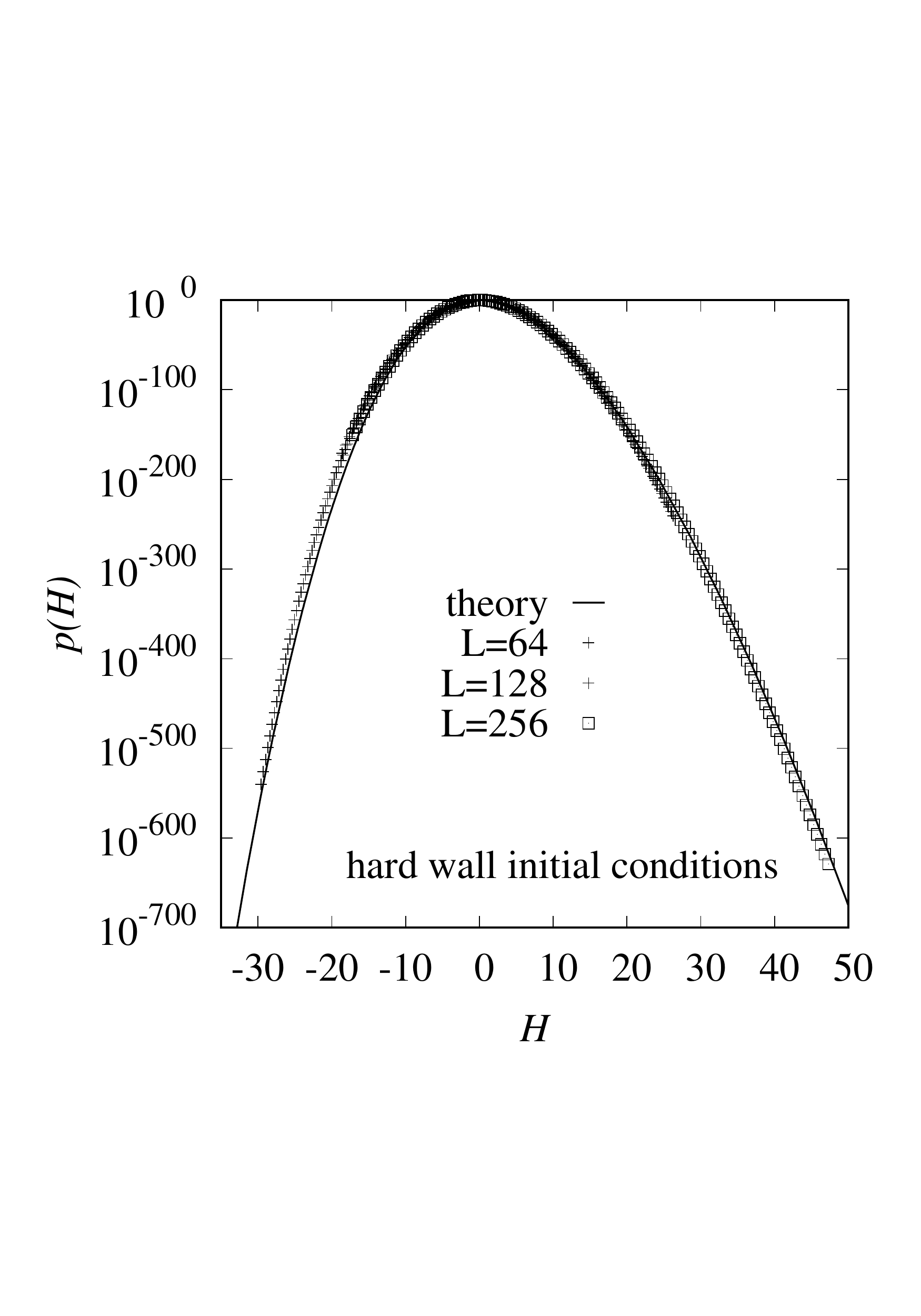}
\caption{  Probability distribution $P(H,t)$ for a short time $t=1/16$ for three
different polymer lengths $L=\lbrace 64,128,256 \rbrace$ for the droplet initial condition in a half-space with an infinite hard-wall at the origin. The solid line indicates the analytical result displayed in 
Eq. \eqref{eq:STdropletHWHW} obtained in \cite{krajenbrink2018large}.
\label{fig:P_H_0625HW1}}
\end{figure}

 The numerical simulations for the droplet initial condition in half-space in the presence of a hard wall were run for polymers of length $L=\lbrace  64,128,256 \rbrace$ and temperature $T$ chosen so that the corresponding time for the Kardar-Parisi-Zhang equation is fixed at $t=1/16$. Convergence to the analytic predictions is expected for $L\to +\infty$. We present the simulations in Figs.~\ref{fig:P_H_0625HW1} and \ref{fig:P_H_0625HW2} and we observe that the agreement between numerical and analytic results is remarkable in the tails (for $H\leqslant -25$ and all $H\geqslant 0$), but nonetheless there is an intermediate region for negative $H$ ($-20\leqslant H \leqslant -10$) where the matching is not entirely perfect: this could possibly come from finite size effects but it still need to be further investigated.

\section{Conclusion and outlook}
To summarize, a large-deviation sampling approach has been used to measure the probability distribution $P(H,t)$ of the centered height $H$ solution the KPZ equation, for various initial conditions : flat and stationary in full-space and droplet in half-space in presence of an infinitely repulsive wall. This was achieved using a lattice directed polymer model, whose free energy converges, in the high-temperature limit, to the height of the continuum KPZ equation. This allowed to determine numerically the probability distribution of the height over a large range of values, leading to a precise comparison with the analytical predictions. We find that the agreement with the short-time large-deviation function $\Phi(H)$ predicted by the theory, which differs according to the geometry and the initial condition, is spectacular, even very far in the tails. The existence and location of a phase transition for the stationary initial condition are fully confirmed, as well as the close connection of the rate function for the flat and stationary initial conditions.

It would be interesting to extend this study in several directions. First, multi-point distributions of the height field could be
investigated. Recently the optimal profile at time $t/2$, i.e. $h(x,t/2)$, conditioned to a large value of
$|H|$ (at time $t$), predicted by the weak noise theory was successfully tested in a similar numerical simulation \cite{hartmann2019optimal}. Although this provides some information, one could ask more detailed questions
about the configurations in real space of the polymer corresponding to
the tails. One expects that the polymer configurations contributing to the left tail 
of the distribution of $H$ will be quite different from the ones which contribute to the right tail. It would be interesting to determine the roughness of these atypical polymers, both at short time
and at large time.  The knowledge of the shape of such polymers could be of great benefits to get some insight on how to sample from atypical polymer distributions.\\

\acknowledgments 
We thank A. Rosso, G. Schehr and S. N. Majumdar for very helpful discussions. AK and PLD acknowledge support from ANR grant ANR-17-CE30-0027-01 RaMaTraF.

\newpage
\appendix

\begin{widetext} 

\bigskip

\bigskip

\begin{large}
\begin{center}

SUPPLEMENTARY MATERIAL

\end{center}
\end{large}

\bigskip

We give some additional technical details of the results of this manuscript.

\section{The Lambert function $W$}
\label{app:lambert}
We introduce the Lambert $W$ function \cite{corless1996lambertw} which we use extensively throughout the Letter. Consider the function defined on $\mathbb{C}$ by $f(z)=ze^z$, the $W$ function is composed of all inverse branches of $f$ so that $W(z e^z)=z$. It does have two real branches, $W_0$ and $W_{-1}$ defined respectively on $[-e^{-1},+\infty[$ and $[-e^{-1},0[$. On their respective domains, $W_0$ is strictly increasing and $W_{-1}$ is strictly decreasing. By differentiation
 of $W(z) e^{W(z)}=z$, one obtains a differential equation valid for all branches of $W(z)$
\begin{equation} \label{derW} 
\frac{\rmd W}{\rmd z}(z)=\frac{W(z)}{z(1+W(z))}
\end{equation}
Concerning their asymptotics, $W_0$ behaves logarithmically for large argument $W_0(z)\simeq_{z\to +\infty} \ln(z)-\ln \ln (z)$ and is linear for small argument  $W_0(z) \simeq_{z \to 0} z-z^2+\mathcal{O}(z^3)$. $W_{-1}$ behaves logarithmically for small argument $W_{-1}(z)\simeq_{z \to 0^-} \ln(-z)-\ln(-\ln(-z))$. Both branches join smoothly at the point $z=-e^{-1}$ and have the value $W(-e^{-1})=-1$. These remarks are summarized on Fig. \ref{fig:Lambert}. More details on the
other branches, $W_k$ for integer $k$, can be found in \cite{corless1996lambertw}.

\begin{figure*}[ht!] 
\begin{center}
\includegraphics[width = 0.5\linewidth]{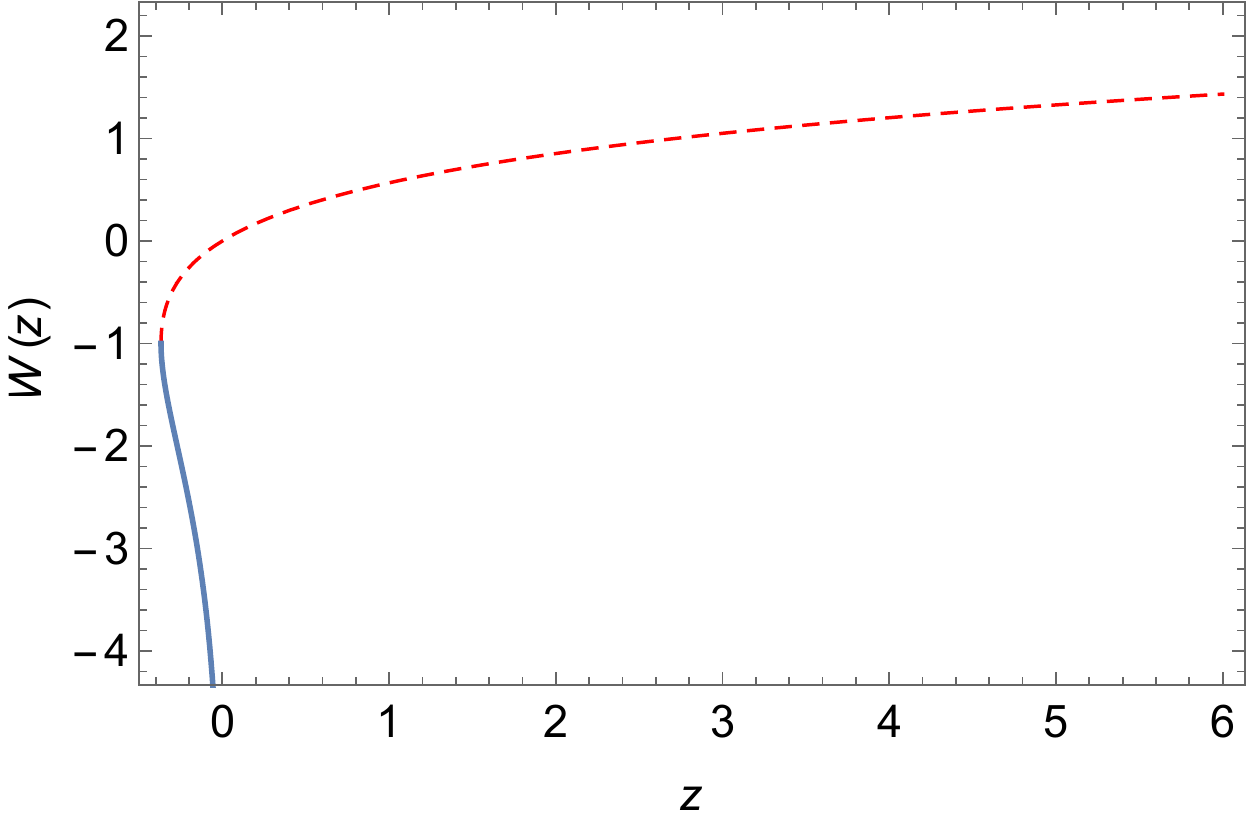}
\caption{The Lambert function $W$. The dashed red line corresponds to the branch $W_0$ whereas the blue line corresponds to the branch $W_{-1}$. }
\label{fig:Lambert}
\end{center}
\end{figure*}
\section{Technical details of the importance sampling algorithm\label{app:sampling}}
 To sample a wide range of values of $H$, one chooses a
suitable set of parameters $\{\theta_{-N_{n}},\theta_{-N_{n}+1},\ldots,
\theta_{N_{p}-1},\theta_{N_{p}}\}$,
$N_{n}$ and $N_{p}$ being the number of negative
and positive parameters, to access the large-deviation regimes (left and right). 
The normalisation constants $Q(\theta)$ are obtained 
by first computing the histogram using simple sampling, i.e., the
bulk of the distribution,
which is well normalised and corresponds to $\theta=0$.
Then for $\theta_{+1}$, one matches the right part of the biased histogram with the left tail of the unbiased one. For a perfect
matching, with a suitable chosen partition function $Q(\theta_{+1})$,
the two distributions would agree in the overlapping region, i.e.,
$e^{\theta H} Q(\theta) P_\theta(H,t) = e^{\theta_{+1} H} Q(\theta_{+1}) 
P_{\theta_{+1}}(H,t)$ $\forall H$ overlapping 
(with $e^{\theta H} Q(\theta)=1$ for $\theta=0$). 
Due to statistical fluctations, this is never
exactly fulfilled, thus we determine $Q(\theta_{+1})$ such that the
squared difference between the two rescaled distribution in the
overlapping regime  is minimized. In the same way
for $\theta_{-1}$ one matches the left part of the biased histogram with the right tail of the unbiased one. Similarly one iterates for the other values of $\theta$ and the corresponding normalisation
constants  can be obtained. Note that in the end the resulting
distribution is slighlty non-normalized, because the tails were added
to the bulk of the distribution, but this is easily repaired
by a final global normalization.

The main drawback of our method is that as for any Markov-chain Monte Carlo simulation, it has to be
equilibrated and this may take a large number of steps. To speed the
simulation up, \emph{parallel tempering}  was used \cite{hukushima1996}. Here, a parallel
implementation using the \emph{Message Passing Interface (MPI)} was
applied, such that each computing core was responsible in parallel 
for an independent realisation
$V_{i}(s)$ at a given $\theta_i$. 
After 1000 Monte Carlo steps, one parallel-tempering
sweep was performed and the parameters  $\theta_i$ and $\theta_{i+1}$
were exchanged between two computing cores. The number and values
of the inverse temperatures are determined by some preliminary test
simulations
with criterion that the empirical
 acceptance rate of the parallel-tempering
exchange steps is about 0.5  for all pairs of $\theta_i, \theta_i+1$ of 
neighboring inverse temperatures. This resulted in few hundred different
values of the inverse temperature where the systems were simulated in 
parallel. For each case considered here 
a corresponding cluster was used for few weeks.
A pedagogical explanation and examples of 
this sampling  procedure can be found in 
Ref.~\cite{align_book}.

\end{widetext}


\begin{thebibliography}{100}
\bibitem{footnote100} This is equivalent to use everywhere the following units of space, time and heights\\
$ x^*=(2 \nu)^3/(D \lambda_0^2), \quad t^*=2(2 \nu)^5/(D^2 \lambda_0^4) , \quad 
h^*=\frac{2 \nu}{\lambda_0}$
\end{thebibliography}
\end{document}